# Topological phononic insulator with robustly pseudospin-dependent transport

Bai-Zhan Xia[1*], Ting-Ting Liu[1], Hong-Qing Dai[1], Jun-Rui Jiao[1], Xian-Guo Zang[2], De-Jie Yu[1*], Sheng-Jie Zheng[1], Jian Liu[1*]

1 State Key Laboratory of Advanced Design and Manufacturing for Vehicle Body, Hunan University, Changsha, Hunan, People's Republic of China, 410082)

2 College of Transportation & Logistics, Central South University of Forestry & Technology, Changsha 410004,Hunan

Topological phononic states, facilitating acoustic unique transports immunizing to defects and disorders, have significantly revolutionized our scientific cognition of acoustic wave systems. Up to now, the theoretical and experimental demonstrations of topologically protected one-way transports with pseudospin states in a phononic crystal beyond the graphene lattice with $C_{6v}$ symmetry are still unexploited. Furthermore, the tunable topological states, in form of robust reconfigurable acoustic pathways, have been evaded in the topological phononic insulators. Here, we realize a topological phase transition in the double Dirac degenerate cone of rotatable triangular phononic crystals with $C_{3v}$ symmetry, by introducing the zone folding mechanism. Along a topological domain wall between two portions of phononic crystals with distinct topological phases, we experimentally observe the quantum spin Hall (QSH) effect for

---

xiabz2013@hnu.edu.cn (Baizhan Xia)

djyu@hnu.edu.cn (Dejie Yu)

liujian@hnu.edu.cn (Jian Liu)

sound, characterized by the robust pseudospin-dependent one-way edge model. As the triangular phononic crystals can freely rotate, we straightforward reconfigure an arbitrary contour defined by the topological domain wall along which the acoustic waves can unimpededly transport. Our research develops a new route for the exploration of the topological phenomena in experiments and provides an excellent framework for freely steering the acoustic immune-backscattering propagation within topological phononic structures.

The intriguing discovery of a novel state of the condensed matter, known as the topological order in the quantum spin Hall (QSH) effect system and the topological insulator[1-4], has inspired researches for analogous states in bosonic systems, such as periodic photonic crystals[5-24]. As the electric transport is driven by potential gradients which are unsuited for the photonic transport, the topological state of photon should be excited in a dissimilar way. External magnetic fields, which could break the time reversal (TR) symmetry of photonic systems, were initially utilized to realize the topological orders of photons[6-10, 21]. At the optical frequencies, materials with strong magnetic responses are extremely absent. Even at the microwave frequencies, the practical application of the strong magnetic field is exceptionally inconvenient. A viable alternative, namely the dynamic modulation of system parameters, has been successfully used to emulate the effects of magnetic fields for the desirable topological properties[11, 12, 25-29]. Recently, the optical bianisotropic metamaterials, supporting the graphene-like topological insulators with the strong spin-orbit interactions analogous to the QSH effects of the condensed matter systems, has been developed [14-16, 30, 31].

As the speed of sound is orders of magnitude less than the velocity of light, the phononic propagation possesses a smaller wavelength and a stronger phonon-phonon interaction[32]. Furthermore, the slow group velocity and the high density of phonon could lead to the strong backscattering from defects and disorders[33]. Topological condense matters feature the robust one-way edge states against backscattering. Thus, in analogy with photons, phonons can also get benefit from the topologically robust states, such as affording the unparalleled tolerances against defects over wide ranges. What's more, the research on the topological orders of phononic states is of scientific significance. Firstly, phonons have three available spin states[34], which essentially differs from electrons and photons with two spin states. Due to the crucial role of the spin state in the formation of the topological insulator, the new degree of freedom of the spin state can realize a new topologically protected acoustic transport. Secondly, in comparison with electrons and photons, topological phononic insulators provide the macroscopic platforms for the investigation of the quantum correlation and the quantum topology. Unfortunately, in the airborne sound, the traditional spin-orbital interaction mechanism is invalid because of the inherent longitudinal nature of the acoustic polarization. Aiming at this issue, scientists have conducted fruitful explorations. For the gyroscopic mechanical systems, the phononic helical edge states against defects and disorders have been achieved in the time-asymmetric gyroscopes emulating "magnetic fields"[35-38]. For the fluid acoustic systems, the topologically protected edge modes have been theoretically realized in networks of acoustic cavities with circulating airflow[39-45] and spatiotemporal modulation[46, 47]. Dynamic instability and inherent noise arising from moving background are detrimental in their engineering applications. The phononic honeycomb lattice with $C_{6v}$ symmetry possesses an accident double Dirac cone with a four-fold degeneracy [48-53]. When the honeycomb lattice undergoes a symmetry inversion, the topological phase transition, inspired by the band inversion near the double Dirac degenerate cone, lead to a robust pseudospin-dependent one-way edge model[54]. Recently, Dirac and Dirac-like cones beyond the honeycomb lattice have been uncovered[55-57]. Based on the group theory, the double Dirac cone cannot appear in the artificial crystal lattice beyond $C_{6v}$ symmetry. Introducing an energy band folding mechanism, two Dirac cones at the K and K' points in the first Brillouin zone of the photonic honeycomb lattice can fold to a double Dirac cone at the $\Gamma$ point[21]. Even though this folded double Dirac cone still emerges in the honeycomb lattice with $C_{6v}$ symmetry, the energy band folding mechanism undoubtedly provides an intriguing way to reveal the new type of the double Dirac cone. If we can determine that the double Dirac denegation cone can be folded in a phononic lattice beyond $C_{6v}$ symmetry, a new class of the topological phononic insulator with various symmetries may be developed. Furthermore, a significant property of topological insulators obtainable in condensed matter systems, namely the well-controllable topological state, has evaded in phononic systems. The well-controllable topological state through reconfigurable synthetic acoustic media would offer great flexibility for manipulating topologically protected acoustic transport and promise wonderful functionality ranging from spin/helicity filtering to tunable phononic devices.

Based on this idea, we experimentally demonstrate that the acoustic wave can robustly transport along the arbitrarily shaped pathway and can be steered to any point within a reconfigurable topological phononic insulator with $C_{3v}$ symmetry. This property is obtained in a phononic lattice consisting of rotatable triangular prisms. We show that a double Dirac cone folded by two Dirac cones emerges in a phononic lattice with $C_{3v}$ symmetry. When the triangular prisms rotate from left to right, the phononic triangular lattice undergoes a spatial symmetry inversion and produced an analogy of the strong spin-orbit coupling, leading to a topological phase transition. Along the domain wall between two portions of prisms with distinct topological phases, we experimentally observe the quantum spin Hall (QSH) effect for sound, characterized by the robust pseudospin-dependent one-way edge transport inside the topological bulk bandgap. As prisms can be freely rotated right or left, we straightforward reconfigure arbitrary sharped pathways along which acoustic waves can robustly transport against defects, such as cavities, disorders, etc.

**Tunable topological phononic insulator**

The most significant property of topological nontrivial states of phononic systems is their unique ability to realize topologically protected phononic propagation in one way [39-47] or spin-polarization[34, 54, 58]. The first key condition for topologically protected transports is to produce a phononic band structure with a double Dirac degenerate cone. Consider a phononic crystal lattice formed by an array of rotatable triangular prisms, illustrated in Fig. 1a. Two types of unit cell and their corresponding Brillouin zones are selected. The first one is a primitive hexagonal unit cell containing a triangular prism (marked by the red line) and the second one is a larger hexagonal unit cell including three triangular prisms (marked by the blue line). The larger hexagonal unit cell is three times larger of the primitive one, so the first Brillouin zones of the larger hexagonal unit cell is a third of the primitive one. The $\Gamma_{II}$-$M_{II}$ of the larger hexagonal unit cell is equivalent to $\Gamma_I$-$M_{II}$, $K_I$-$M_I$ and $K'_I$-$M_{II}$ (green lines) of the primitive hexagonal unit cell. When taking the primitive hexagonal unit cell, this phononic crystal with $C_{3v}$ symmetry carries the Dirac dispersions at $K$ and $K'$ points in the first Brillouin zone, shown in Fig. 1b. According to the group theory, these Dirac degenerate cones are deterministic ones, independent of the filling ratio[55]. When taking the larger hexagonal unit cell, this phononic crystal carries a double Dirac dispersion at the $\Gamma$ point in the first Brillouin zone, illustrated in Fig. 1c. The four-fold degeneracy at the double Dirac cone possesses two types of phononic modes, classed as the symmetric (*S*) mode and the anti-symmetric (*A*) mode (illustrated in Fig. 1c). Both the *S* and *A* modes are originated from two degenerate Broch modes of Dirac cones. Thus, this double Dirac degenerate cone is nothing but a folding of two Dirac cones at the *K* and *K'* points. Matching frequencies of the *S* and *A* modes over the extend frequency band, these phononic modes are referred to polarization or spin-degeneration[16, 30, 31, 34, 54], by emulating two electronic spin states arising from QSH effects[1-4].

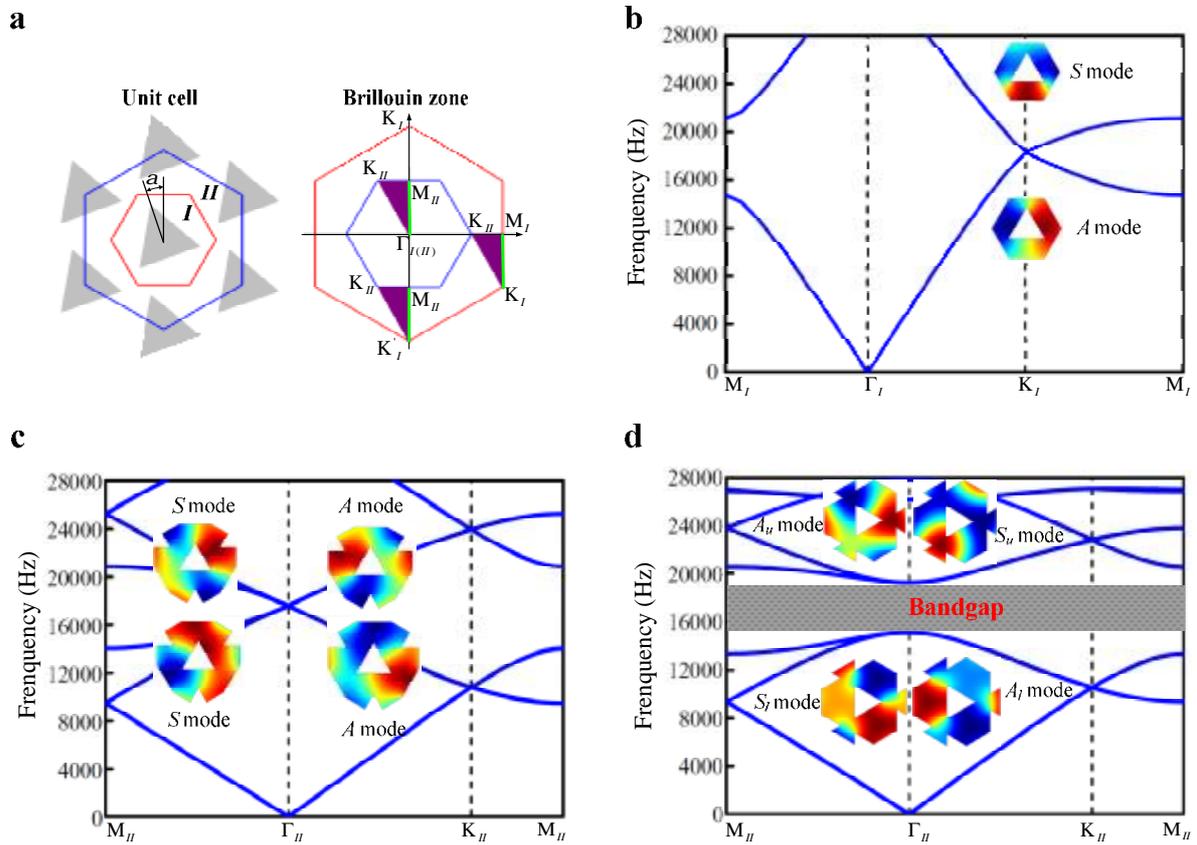

**Figure 1| Schematic of phononic crystal and its band structures. a**, Left panel: a geometric arrangement of our phononic crystal consisted of rotatable triangular prisms with a lattice constant 10mm. The length of the triangular prism is 6.8mm. Right panel: Brillouin zones of the primitive and larger hexagonal unit cells. The density and longitudinal sound speed are 7,800 kg/m$^3$ and 6,010m/s for stainless steel and 1.25 kg/m$^3$ and 343m/s for air. **b**, Bulk band structure with a Dirac degenerate cone at the *K* point in the Brillouin zone of the primitive hexagonal unit cell. Pressure field distributions at the Dirac degenerate cone are inserted in the band structure. **c**, Bulk band structure with a double Dirac degenerate cone at the *Γ* point in the Brillouin zone of the larger hexagonal unit cell. Pressure field distributions at the double Dirac degenerate cone are inserted in the

band structure. **d**, Bulk band structure with a complete bandgap induced by left rotating triangular prisms by 30 degrees. Pressure field distributions in the vicinity of the double Dirac degenerate cone are inserted in the band structure.

The second critical condition for a topological state is the topological phase transition accompanied with the opening of the double Dirac degenerate points[30, 31, 34]. As this folded double Dirac cone is induced by the $C_{3v}$ symmetry of the primitive hexagonal unit cell, its degeneracy cannot be lifted by only tuning its filling ratio (illustrated in Fig. S1 in the Supplementary Information). A viable alternative to lift the degeneracy of the folded double Dirac cone is to reduce the inversion symmetry which plays a significant role in the degeneration of the Dirac cones at the *K* and *K'* points. The angular dependent frequencies for the band edges near the Dirac degenerate cone at the *K* point and the double Dirac degenerate cone at the *Γ*-point are illustrated in Fig. S2 in the Supplementary Information. Rotating prisms left by 30 degrees (shown in Fig. 1d), a synthetic gauge field with a reduced inversion symmetry to emulate the spin-orbit coupling is yielded. This synthetic gauge field opens a complete topological phononic bandgap in the vicinity of the original double Dirac cone. This topological bandgap in the bulk crystal forms an "insulating states". The reduction of the inversion symmetry induces a pairwise coupling between the original Dirac bands: the upper symmetric ($S_u$) mode and the lower anti-symmetric ($A_l$) mode, as well as the upper anti-symmetric ($A_u$) mode and the lower symmetric ($S_l$) mode (illustrated in Fig. 1d). For all frequencies in the proximity of the original double Dirac cone, these hybridized system eigenmodes can efficiently emulate the new pseudo-up-spin and pseudo-down-spin[31, 34, 54].

When prisms rotate from left (or right) with an angle 30 degrees to right (or left) with an angle 30 degrees, a band inversion takes place (illustrated in Fig. S3 in the Supplementary Information). This confirms a topological phase transition between prisms rotating right and left by 30 degrees.

The last important condition for a topological phononic insulator is the produce of the helical edge state without causing coupling of two spin states. For this purpose, we configure a boundary across which the topological phases of phononic crystals are opposite. As shown in Fig. 2a, this boundary is a domain wall between two areas of prisms whose rotation directions are opposite: the upper part rotating right by 30 degrees and the lower part rotating left by 30 degrees. According to the bulk-boundary correspondence principle[59], a pair of edge states (red and blue lines) emerge inside the overlapped bulk bandgap of two distinct phononic crystals (Fig. 2b). The topological edge modes characterize the topological helical states analogous to the unidirectional spin-polarized one-way propagation in the condensed matter systems[16, 30, 31, 34, 54]. Therefore, although the symmetry inversion yields a topological phononic bandgap in which the acoustic propagation is efficiently prevented, the spin of the edge states localized at the interface of two distinct phononic crystals leads to a topologically protected one-way edge model, ensuring the pseudospin-dependent one-way propagation.

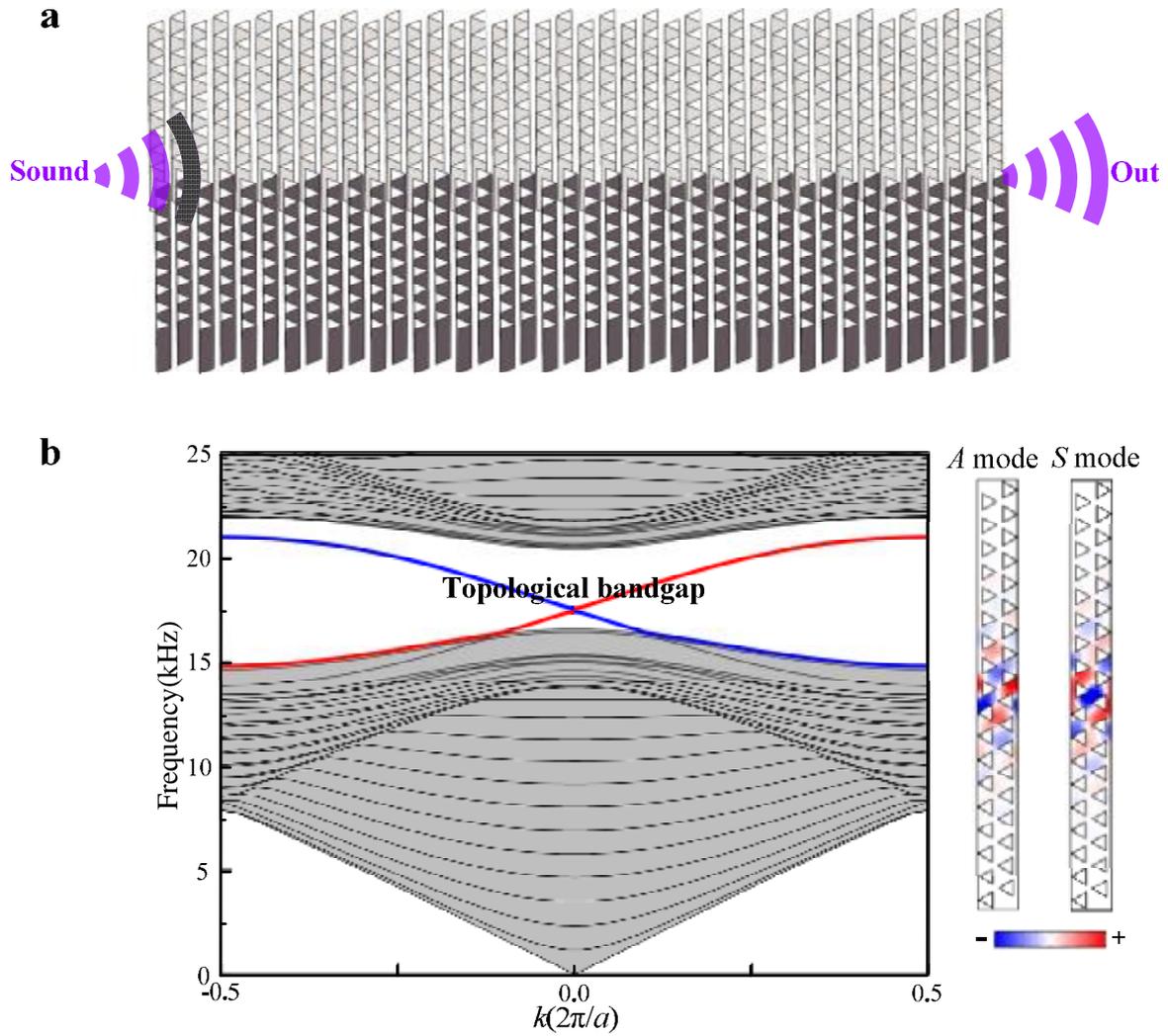

**Figure 2| Reconfigurable topological phononic insulator and its edge band structures. a**, a schematic of our topological phononic insulator with a bianisotropic domain wall across which the triangular prisms oppositely rotate. **b**, Left panel: phononic band structure of a supercell consisting of 18×2 topological phononic crystals. The blue and red lines respectively represent acoustic spin+ and spin- edge states which are hybridized with a symmetric edge mode (S) and anti-symmetric edge mode (A). The grey lines indicate the bulk bands. The pale green shaded region shows the topological bandgap. Right panel: pressure field distributions of the supercell at $k=0$ for the S and A modes.

**Pseudospin-dependent edge mode of topological phononic insulator**

Generally, it was considered as an extremely challenging work to observe a spin phenomenon in experiment. Here, we utilized a cross-waveguide splitter[30, 60], which has been used to study the photonic pseudospin transport with a very high fidelity even when the pseudospin-dependent state is unknown. As shown in Fig. 3a, the splitter is divided into four sections with four input/output ports (marked as 1, 2, 3 and 4). In the top-left and down-right sections, the triangular prisms of the topological phononic insulator rotate left by 30 degrees. In the top-right and down-left sections, the triangular prisms rotate right by 30 degrees. For the acoustic states, when the acoustic wave propagates from port 1 (or port 3) to port 2 and port 4 by crossing the junction, the triangular prisms rotating left and right are respectively located on the left and right sides of these sound propagation paths. As the pseudospin-dependent state is determined by the spatial symmetry, the pseudospin-dependent edge states of these sound propagation paths with same spatial symmetries are always permitted. In this case, the acoustic wave from the port 1 (or port 3) can propagate through the junction to the port 2 and port 4 (labelled by purple lines). Similarly, the structural spatial symmetries of the sound propagation paths from port 2 (or port 4) to port 1 and port 3 are also preserved. Thus, the acoustic wave can transport along these sound propagation paths (labelled by green lines). However, for the sound propagation path from port 1 to port 3, the rotation directions of triangular prisms located on both sides of the path are counterchanged when crossing the junction. Due to the inversion of the structural spatial symmetry, the pseudospin states crossing the junction are also inversed. These opposite pseudospin states cannot be excited by each other due to the mismatch of their spin configurations, indicating that the acoustic wave cannot propagate from port 1 to port 3.

The similar results can be obtained for other straight through fashions (from port 3 to port 1, from port 2 to port 4 and from port 4 to port 2). Therefore, we can come to a conclusion that because of the opposite slope of the dispersion band for each individual acoustic spin edge state in Fig 3a, the pseudospin-dependent edge state can only support a topological one-way propagation with a anticlockwise (purple circular arrows) or clockwise (green circular arrows) direction, as shown in Fig. 3a. These conclusions are also perfectly agreement with the simulation results illustrated in Fig. 3b-c and the experimental measures presented in Fig. 3d-e. Such a topological pseudospin behavior, characterizing an acoustic counterpart of the QSHE, cannot normally appear in the non-topological structure. For a conventional phononic crystal with a cross-waveguide, the acoustic wave cannot propagate from any port to the other three ports because of its reciprocal phase shift characteristics and the Bragg scattering from triangular prisms (see Fig. S4 in the Supplementary Information). On the contrary, the acoustic propagation in our cross-waveguide splitter is non-reciprocal due to the pseudospin property of the topological phononic insulator, which leads to a broadband pseudospin-dependent wave splitting.

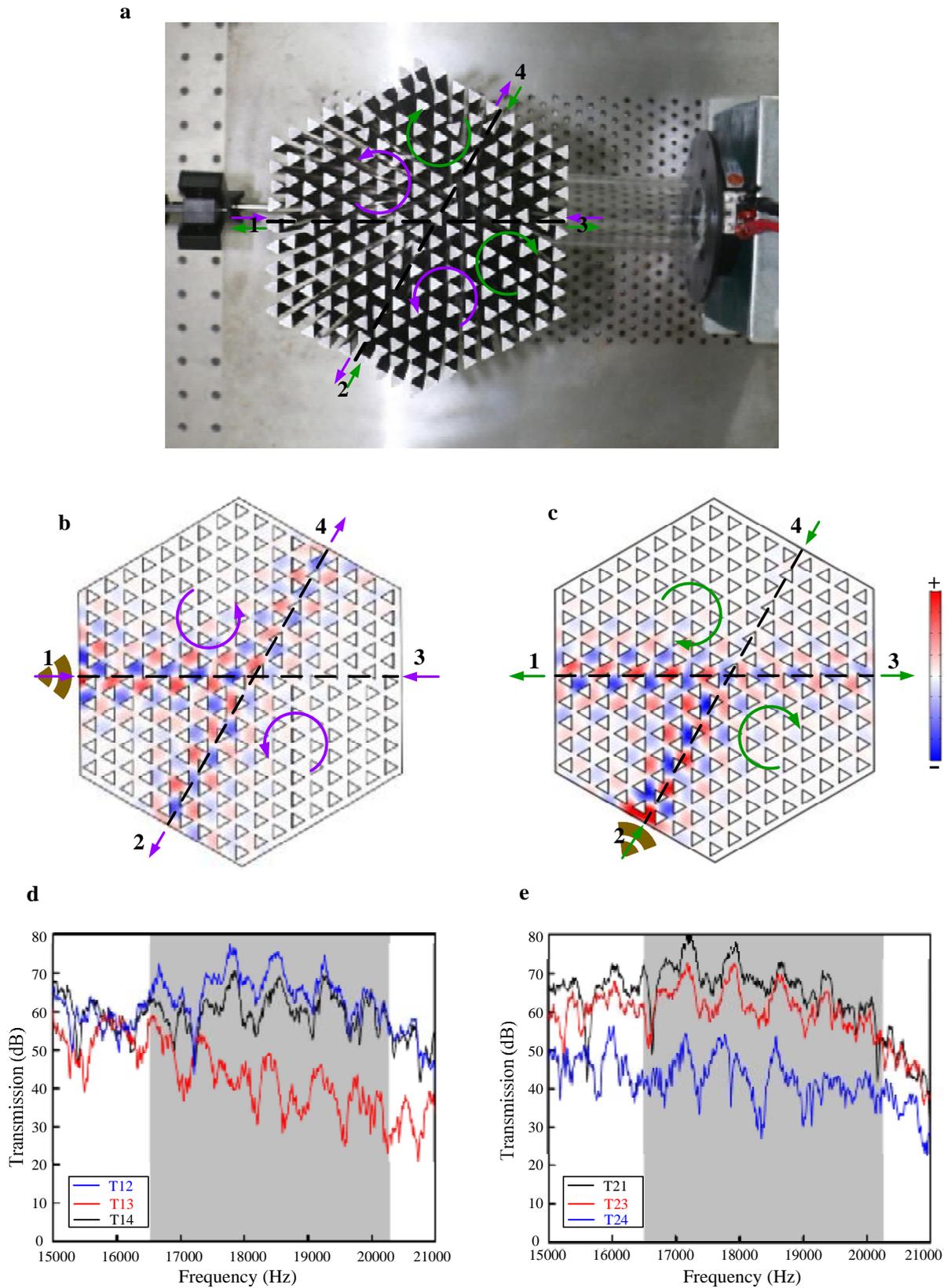

**Figure 3 | Acoustic pseudospin-dependent edge mode at our topological cross-waveguide splitter**. **a**, The

photo of our topological cross-waveguide splitter. In this test, the anticlockwise (clockwise) edge circulating propagation is indicated by the purple (green) circular arrows. **b**, **c**, Simulated acoustic pressure field distributions at a frequency of 18.42 kHz (within the bulk bandgap) for the cases that the acoustic waves are incident from port 1 and port 2. **d**, **e**, Experimental transmission spectra for the acoustic waves are incident from port 1 and port 2. $T_{ij}$ indicates the transmission spectrum from port $i$ to port $j$ ($i$, $j$=1, 2, 3,4). The shadow regions indicate the topological bandgaps.

**Topologically protected acoustic propagation against defects**

The key feature of the topological phononic insulator is its topologically robust one-way edge state against backscattering from defects or disorders. To verify the robust transport property of our topological phononic insulator, two cases with different defects were intentionally introduced into the cross-waveguide splitter. The first one is an incomplete cross-waveguide splitter with several cavities (rounded by green circles). The other one is a cross-waveguide splitter with several disordered prisms which inversely rotate (rounded by green circles). Multiple resonance dips arising from the cavity resonances will result in a significant perturbation on the acoustic propagation in the bandgap-guiding waveguide of trivial phononic crystals[54]. Strong disorders are well known to the localizing and backscattering characteristics in the bandgap-guiding waveguide without topological protection[54]. Cavities and disorders are not the spin-mixing defects, indicating that the topological characteristics of our phononic insulator will not be broken by these "nonmagnetic" impurities[9, 34, 54, 61]. Stimulated acoustic pressure fields, illustrated in Figs. 4a-b, show that the acoustic waves from the port 1 can effectively detour cavities and disorders, transmitting to the ports 2 and 4. Experimental transmission spectra, plotted in Fig. 4c-d, also indicate that the acoustic wave can exhibit a robustly pseudospin-dependent propagation in the topological cross-waveguide splitter with "nonmagnetic" defects. These simulation and experimental results expectantly verify the topologically protected transport property of our phononic insulator.

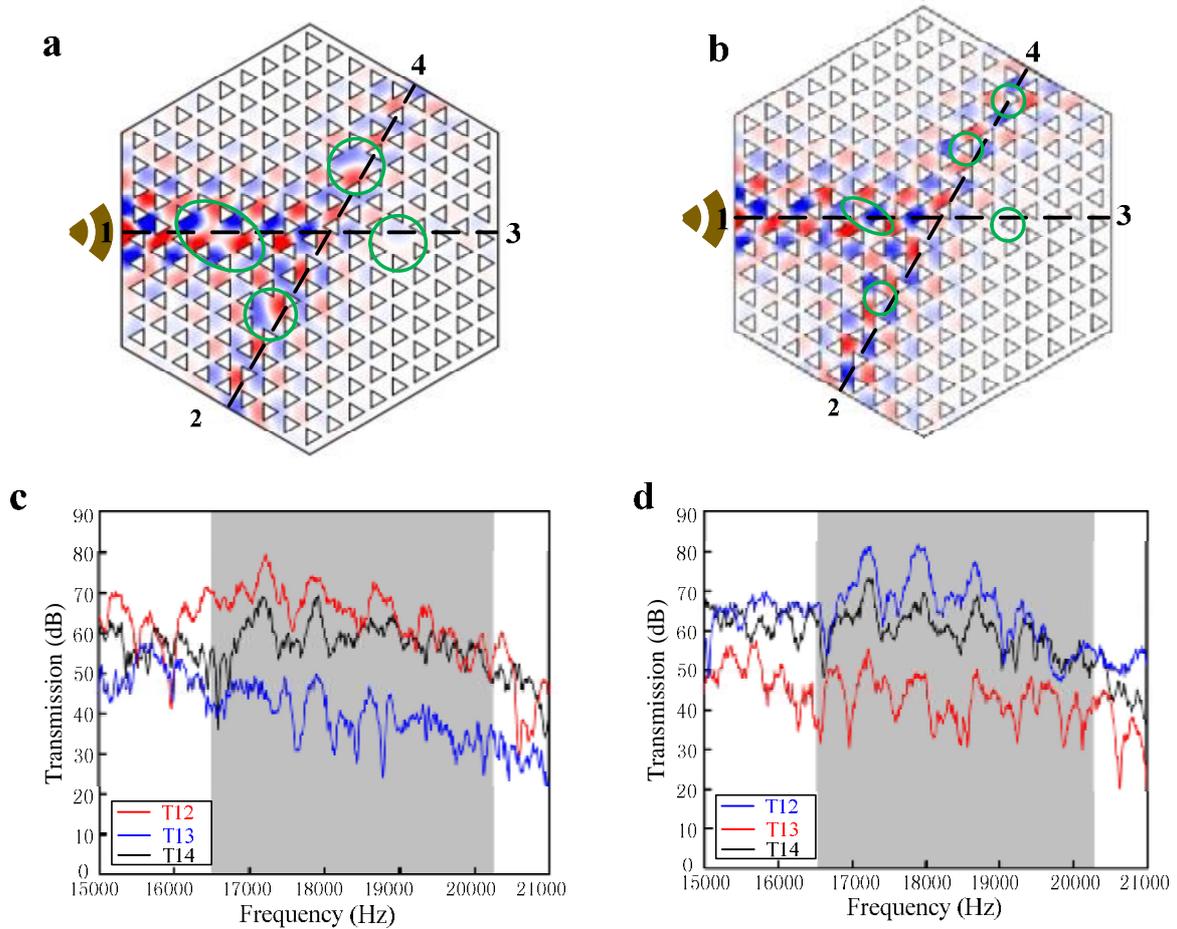

**Figure 4 | Robust one-way acoustic transport. a-b**, Stimulated acoustic pressure field distribution in the topological cross-waveguide splitter with cavities and disorders. The frequency of the acoustic wave is 18.42 kHz (within the bulk bandgap). **c-d**, Experimental transmission spectra of the topological cross-waveguide splitter with cavities and disorders. The acoustic waves are incident from port 1. $T_{12}$, $T_{13}$ and $T_{14}$ indicate the transmission spectra from port 1 to ports 2, 3 and 4. The shadow regions indicate the topological bandgaps.

**Reconfigurable guiding of the topological edge mode.**

Prisms in our topological phononic insulator can freely rotate around their central axes. By rotating some prisms in the left direction and the other ones in the right direction, arbitrarily sharped contours between distinct phononic crystals with opposite topological phases can be created. Thus, in contrast to passive topological phononic insulators with fixed phononic crystals[34, 54], our topological phononic insulator not only expresses a new avenue to realize the topological state, but also provides a well-controllable manner to freely reconfigure robust acoustic pathways. To confirm this robust reconfigurable transport property, two nodes (A3 and A4) respectively locating at the down-left and upper-right regions are selected. Fig. 5a shows that the acoustic wave incident from the input port 1 effectively passes through the node A3 and the node A4, and finally reaches output port 2 with a high transmission. Fig. 5c shows that the acoustic wave incident from the input port 1 effectively passes through the node A4 and the node A3, and finally reaches output port 2 with a high transmission. As shown Fig. 5b and d, compared with the measured spectra of transmission through the ordinary phononic crystals, the above two configurations of topological phononic insulators exhibit a much higher transmission over the frequency range of the bandgap. The other complicated pathways, including an asteroid pathway and a "HNU"-like pathway, are illustrated in Fig. S5 in the Supplementary Information. These results show that within our topological phononic insulator, the acoustic wave can effectively travel along the arbitrary pathway defined by the domain wall between two portions of phononic crystals with distinct topological phases. This unparalleled ability to freely steer the robust topological propagation opens up a wonderful approach to design tunable topological acoustic devices.

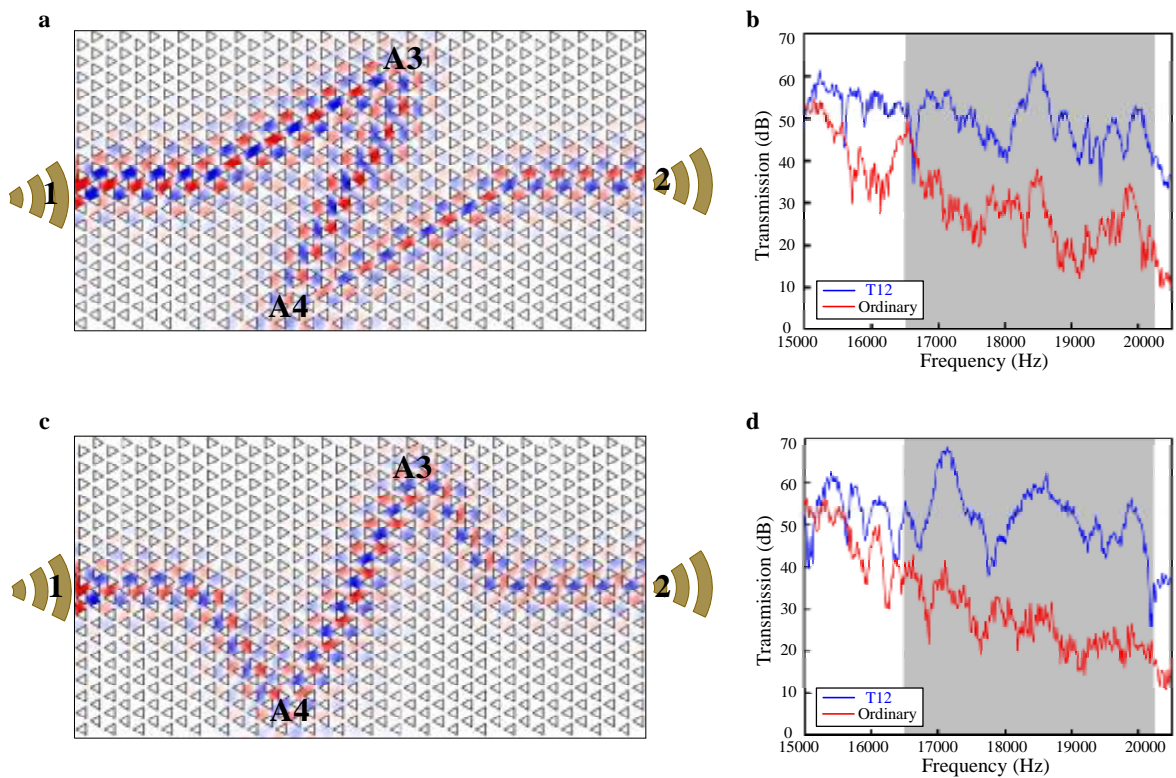

**Figure 5 | Reconfigurable guiding of the topological edge modes. a**, Stimulated acoustic pressure field distribution along the pathway through nodes A3 and A4. **c**, Stimulated acoustic pressure field distribution along the pathway through nodes A4 and A3. Reconfigurable pathways are domain walls between two crystal structures with opposite topological phases arising from the inversion of their rotational directions. **b, d**, Experimental transmission spectra for two acoustic pathways. The blue curve corresponds to the transmission spectra in topological phononic insulators. The red curve corresponds to the transmission spectra in ordinary phononic crystals. The shadow regions correspond to the topological bandgaps.

**Summary and outlook**

To conclusion, we develop a reconfigurable topological phononic insulator without inherent noises arising from dynamical modulation. Our topological platform theoretically and experimentally realizes the fascinating phononic pseudospin phenomenon with a very high fidelity by introducing a cross-waveguide splitter, and exhibit the topologically protected acoustic transports immunizing to "nonmagnetic" defects, such as cavities and disorders. In addition, the robust reconfigurable one-way edge state with the controllable contour in our topological phononic insulator opens up infinite possibilities for manipulating and steering the acoustic wave along arbitrary pathway to any desired point without backscattering. Note that our reconfigurable topological phononic insulator can be easily extended to a broad acoustic spectrum from audible sound to ultrasound, and even up to hypersound, by modulating the lattice constants and the filling rates of triangular prisms. More importantly, the intriguing topological phenomenon of our reconfigurable phononic insulator induced by a folded double Dirac cone of the phononic lattice with $C_{3v}$ symmetry is a milestone in the design of modern topological pseudospin phononic devices beyond $C_{6v}$ symmetry, and generate an impressive potential for applications in a foreseeable future.

**Acknowledgments**

The paper is supported by National Natural Science Foundation of China (No.11402083, 11572121), Collaborative Innovation Center of Intelligent New Energy Vehicle and the Hunan Collaborative Innovation Center of Green Automobile.